\documentclass[fleqn,usenatbib]{mnras}
\usepackage{graphicx}
\usepackage{graphics}
\usepackage{dirtytalk}
\usepackage{txfonts}
\usepackage{natbib}
\usepackage[T1]{fontenc}
\bibpunct{(}{)}{;}{a}{}{,}

\title[The true nature of hard X-ray sources]{Investigating the true nature of three hard X-ray sources}

   \author[M. Molina et al.]{M. Molina$^{1}$\thanks{E-email: manuela.molina@inaf.it (MM)}
          A. Malizia$^{1}$
          N. Masetti$^{1,3}$
          L. Bassani$^{1}$
          A. Bazzano$^{2}$
           A.J. Bird$^{4}$
          M. Fiocchi$^{2}$
          E. Palazzi$^{1}$
          \newauthor
          P. Ubertini$^{2}$
\\          
$^{1}$OAS/INAF, ia Gobetti 101, I-40129 Bologna, Italy\\         
$^{2}$IAPS/INAF, via Fosso del Cavaliere 100, I-00133 Rome, Italy\\
$^{3}$ Departamento de Ciencias F\'isicas, Universidad Andr\'es Bello, Fern\'andez Concha 700, Las Condes, Santiago, Chile\\
$^{4}$Department of Physics and Astronomy, University of Southampton, Highfield, Southampton SO17 1BJ, UK
}

   \date{Accepted XXX. Received YYY; in original form ZZZ}
   \pubyear{2021}

\begin{document}
\label{firstpage}
\pagerange{\pageref{firstpage}--\pageref{lastpage}}
\maketitle

\begin{abstract}
Many of the new high energy sources discovered both by INTEGRAL/IBIS and  Swift/BAT have been characterised thanks 
to extensive, multi-band follow-up campaigns, but there are still objects whose nature remains to be asserted. In this paper we 
investigate the true nature of three high energy sources, IGR J12134-6015, IGR J16058-7253 and Swift J2037.2+4151,
employing multiwavelength data from the NIR to the X-rays.
Through Gaia and ESO-VLT measurements and through Swift/XRT X-ray spectral analysis, we re-evaluate the classification for 
IGR J12134-6015, arguing that the source is a Galactic object and in particular a Cataclysmic Variable.
We were able to confirm, thanks to NuSTAR observations, that the
hard X-ray emission detected by INTEGRAL/IBIS and Swift/BAT from IGR J16058-7253 is  coming from two Seyfert 2 galaxies
which are both counterparts for this source. Through optical and X-ray spectral analysis of Swift J2037.2+4151 
we find that this source is likely part of the rare and peculiar class of Symbiotic X-ray binaries and displays 
flux and spectral variability as well as interesting spectral features, such as a blending of several emission lines
around the iron line complex.
\end{abstract}


\section{Introduction}
In the past 20 years, our knowledge of the high energy sky has greatly improved thanks to
missions such as INTEGRAL \citep{Winkler:2003} and the Neil Gehrels Swift Observatory
\citep{gehrels04}, that with their two main detectors, IBIS \citep{Ubertini:2003} and BAT 
\citep{barthelmy05}, have been continuously monitoring the sky above 14 keV. Both
missions have produced several catalogues, populated by both known sources and objects that are completely
new high energy emitters and whose nature is yet to be determined.
Many of these new hard X-ray sources are common between the INTEGRAL/IBIS  (e.g. \citealt{bird06,bird07,bird10,bird16}) and the Swift/BAT catalogues (e.g. \citealt{baumgartner10,baumgartner13,Oh18}).
In particular in all the INTEGRAL/IBIS surveys released so far  (\citealt{bird16} and \citealt{krivonos21} and 
references therein), about 20\% of the detected high energy sources were completely new discoveries with no clear 
counterpart and no firm classification. 

In the past years, many of these sources have been characterised through an extensive follow-up campaign, 
both in the optical and in the X-ray band resulting in many papers published since 2004 
(see e.g. \citealt{masetti13} and \citealt{landi17}). The variety of sources thus uncovered is truly remarkable, 
spanning from X-ray binaries like Low Mass X-ray Binaries (LMXB, see \citealt{sazonov20} for a review) and
High Mass X-ray Binaries (HMXB, see \citealt{kretschmar19} for a review), 
Cataclysmic Variables (CVs, see \citealt{lutovinov20} for a review) and Symbiotic X-ray Binaries (e.g. 
\citealt{masetti07})  to extragalactic objects, mainly Active Galactic Nuclei (AGN; see \citealt{malizia20} for a 
review). 

Follow-up campaigns of new IBIS and BAT sources have been going on for many years now,
but they still reserve a few surprises, as peculiar 
sources are still found today, several years after the publication of the first IBIS and BAT catalogues (e.g. \citealt{masetti07}). 
Follow-up campaigns at different wavelengths are not only paramount in determining the true nature of new 
discoveries, but are also essential to resolve ambiguities in source identification, as is the case for the 
sources presented in this paper. In this work we aim at 
unveiling the true nature of these three sources, IGR J12134-6015, IGR J16058-7253 and Swift 
J2037.2+4151, for which their classification is still unclear. IGR J12134-6015 and
Swift J2037.2+4151 have both been classified as beamed AGN in several BAT catalogues, but left unclassified in 
various INTEGRAL surveys, while through a multi-band approach we propose a different classification. As
for the third source analysed here, IGR J16058-7253, the issue arises not in the classification of this object as
an AGN, but rather on the origin of the high energy emission,  this object being a blending of two different 
active galaxies. Here we collect all previous and new X-ray data (Swift/XRT, Chandra and NuSTAR) and re-analyse 
them adding the available information at other wavelengths.

\section{X-ray data reduction}

\begin{table*}
\begin{center}
\caption{Observation log of the sources analysed here.}
\vspace{0.2cm}
\resizebox{\linewidth}{!}{%
\begin{tabular}{ccccccc}
\hline
{\bf Source} & {\bf RA}                & {\bf Dec}    &  {\bf Pos. error}  &{\bf Telescope}      &{\bf Obs. date } & {\bf Exp. (ksec)}    \\
           \hline
IGR J12134-6015$^{\dagger}$ &12$^{\rm h}$13$^{\rm m}$24.00$^{\rm s}$& -60$^{\rm d}$15$^{\rm m}$16.541$^{\rm s}$& 0.64\arcsec &Chandra (HRC-I) & 26/02/2011&1.17 \\
                &                                       &                                          & &XRT & 06/02/2011 & 2.86  \\
                &                                       &                                          & &XRT & 09/02/2011 & 3.78 \\
                &                                       &                                          & &XRT & 11/02/2011 & 2.5  \\

                \hline
IGR J16058-7253$^{\ddagger}$ &16$^{\rm h}$05$^{\rm m}$22.8$^{\rm s}$&-72$^{\rm d}$53$^{\rm m}$55.3$^{\rm s}$& 3.8\arcsec& XRT & 15/04/2009& 2.23 \\
(LEDA 259433) &                               &                                        & &XRT & 31/12/2009 & 7.28  \\
                  &                               &                                        & &XRT  & 10/04/2010& 5.14  \\
                  &                               &                                        & &XRT   & 11/05/2010 & 1.73  \\
                  &                               &                                         &&XRT   & 12/05/2010 & 0.66  \\
                  &                                &                                        & &NuSTAR & 01/03/2019 & 22.2  \\
IGR J16058-7253$^{\ddagger}$  &16$^{\rm h}$06$^{\rm m}$06.7$^{\rm s}$&-72$^{\rm d}$52$^{\rm m}$40.6$^{\rm s}$&4.1\arcsec&XRT & 15/04/2009& 2.23 \\
(LEDA 259580) &                                &                                       && XRT & 31/12/2009 & 7.28  \\
                  &                               &                                       & & XRT  & 10/04/2010& 5.14  \\
                  &                               &                                       & & XRT   & 11/05/2010 & 1.73  \\
                  &                               &                                        & &XRT   & 12/05/2010 & 0.66  \\
                  &                                &                                      &  & NuSTAR & 01/03/2019 & 22.2  \\
\hline
Swift J2037.2+4151$^{\ddagger}$&20$^{\rm h}$37$^{\rm m}$05.5$^{\rm s}$&41$^{\rm d}$50$^{\rm m}$05$^{\rm s}$&3.5\arcsec&XRT & 17/08/2006& 5.39 \\
                  &   &&& XRT& 17/12/2006& 4.87 \\
                  &   &&&XRT& 04/12/2007& 2.20  \\

\hline
\end{tabular}
}
\item Notes: $^{\dagger}$: Chandra coordinates; $^{\ddagger}$: XRT coordinates
\label{obs_log}
\end{center}
\end{table*}

Swift/XRT observed all the sources studied here; these data have been previously presented  by \citet{landi11},
however we re-analysed them together with other multiwavelength gathered specifically for this work.
Swift/XRT data reduction was performed using the standard data pipeline
package (\texttt{XRTPIPELINE} v. 0.13.2) in order to produce screened
event files (see \citealt{landi10}). Source events were extracted 
within a circular region with a radius of 20 pixels (1 pixel corresponding to 2.36 arcsec) 
centred on the source position, while background events were extracted from a source-free
region close to the X-ray source of interest. The spectra were
obtained from the corresponding event files using the \texttt{XSELECT}
v. 2.4c software.
We used version v.014 of the response matrices and created 
individual ancillary response files using the task \texttt{xrtmkarf} v.0.6.3.

Chandra HRC-I data reduction for IGR J12134-6015 was performed using \texttt{CIAO-4.13} and \texttt{CALDB\_4.9.4}; the source
coordinates have been obtained with the task \texttt{wavdetect}, which returns the source position; as for
the positional uncertainty, we assume the nominal one of 0.64 arcsec as done by \citet{karasev12}.

NuSTAR data for IGR J16058-7253 were reduced using the \texttt{nustardas\_01Apr120\_V1.9.2} 
and \texttt{CALDB} version 20200429. Spectral extraction and the subsequent production of
response and ancillary files was performed using the
\texttt{nuproducts} task with an extraction radius of 50$^{\prime\prime}$; 
to maximise the signal-to-noise ratio (S/N), the background spectrum was extracted from a
70$^{\prime\prime}$ radius circular region as close to the source as
possible. 

All spectra were binned with \texttt{grppha} in order to achieve a minimum of 20 counts per bin, in order to 
apply the $\chi^2$ statistics and spectral fitting was performed in \texttt{XSPEC} v12.11.1 
\citep{arnaud96}; uncertainties are listed at the 90\% confidence level
($\Delta\chi^2$=2.71 for one parameter of interest). Abundances were all set to Solar and 
the cross-sections employed are photoelectric ones.

\section{IGR J12134-6015}

IGR J12134-6015 was first reported as a high energy emitting source by \citet{krivonos10}, in their 7-year
INTEGRAL all sky survey; these authors suggested that IGR J12134-6015 is associated with the ROSAT source 1RXS 
J121324.5-601458. The source was also listed in the  BAT 58-month and 70-month catalogue 
\citep{baumgartner10,baumgartner13}, where its likely counterpart was identified as the same ROSAT source, 
but no firm identification was reported.
The BAT 105-month \citep{Oh18} and the latest BAT 157-month (\url{https://swift.gsfc.nasa.gov/results/bs157mon/})
catalogues instead classified this source as a beamed AGN, as also did \citealt{krivonos12} in their 9-year 
INTEGRAL survey, whereas in \citet{bird16} 
the source is detected in a 1605.1 day outburst starting on MJD 53292.6, at a 6.3 $\sigma$ level (maximum significance during the outburst), suggesting 
variability at high energy but no classification is given.

The first X-ray follow-up of IGR J12134-6015 was carried out by both \citet{landi11} and \citet{karasev12}, 
who identified the 2MASS counterpart and 
confirmed the association with the ROSAT source; \citet{landi11} also found, 
coincident with this object, the XMM Slew source XMMSL1 J121323.5–601517. 
Spectral properties derived from Swift/XRT data by \citet{landi11}
pointed to a Galactic nature for this source, whereas 
\citet{karasev12} suggested that the source might be an extragalactic object, 
raising the question of what its true nature might be.

The extragalactic nature of this object is also challenged in a paper by \citet{paliya19}, 
where the authors study the physical properties of blazars extracted from the BAT 105-month catalogue;
however, the authors exclude IGR J12134-6015 from the list of BAT blazars,
on the basis of its broad-band properties, which are not typical of beamed AGN.

\begin{figure}
\centering
\includegraphics[scale=0.5]{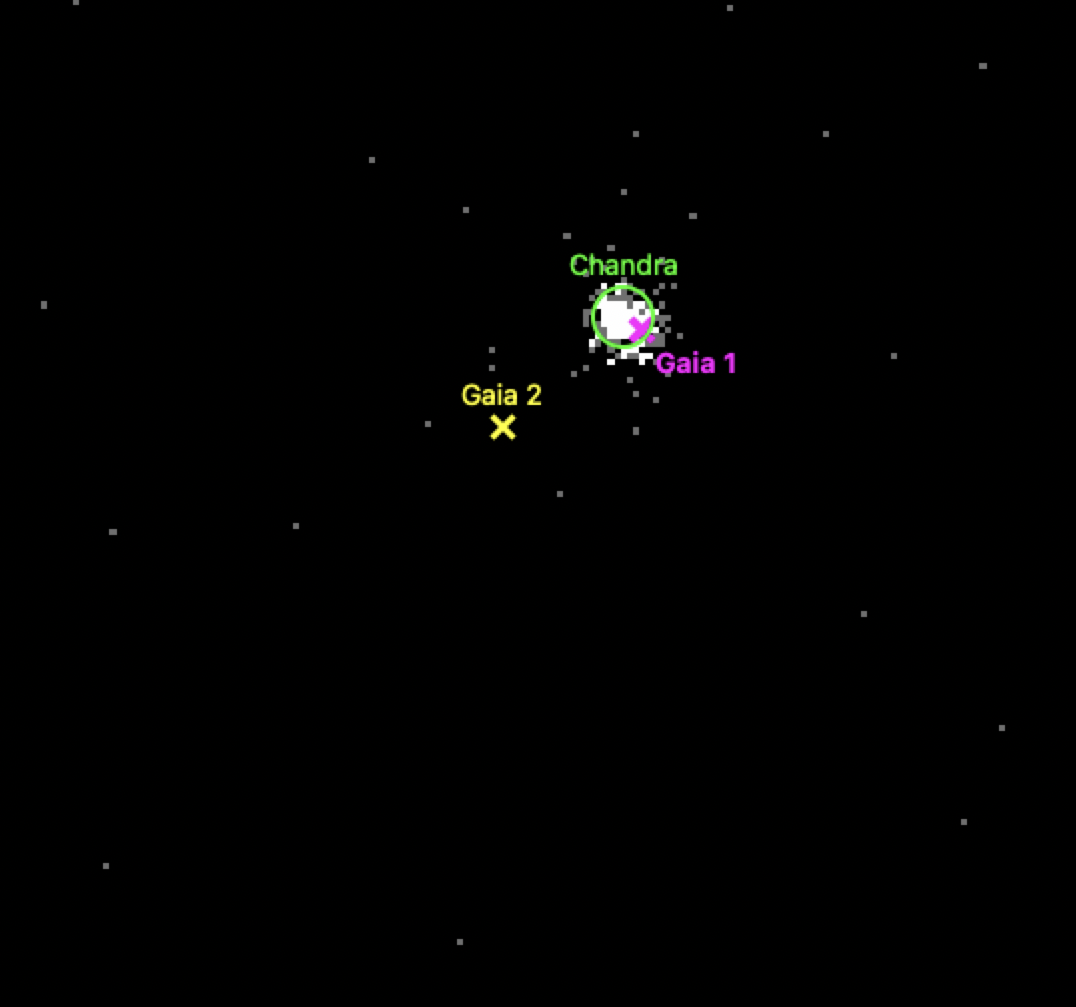}
\caption{{\small Chandra 0.8-10 keV HRC-I image of the sky region around IGR J12134-6015. The green circle is centered on the Chandra coordinates and represents their positional error of 0.64 arcsec (see \citealt{karasev12}). The magenta cross is the likely GAIA 
counterpart of
the X-ray source, while the yellow cross is the second nearby source found in the GAIA catalogue.}}
\label{chandra}
\end{figure}

In order to reach firmer conclusions on the true nature of IGR J12134-6015, we investigated further this
source at multiple wavelengths, starting with all the available X-ray measurements. 
Apart from the Swift/XRT data analysed by \citet{landi11}, IGR J12134-6015 has been observed once by the HRC-I 
instrument on-board Chandra  (see Table \ref{obs_log}).
Analysis of the 0.8-10 keV full band image (see Fig. \ref{chandra})
confirms that there is only one X-ray source 
in the field of view, at the coordinates reported in Table \ref{obs_log} and consistent with the position of the 
ROSAT counterpart. 
Within 4 arcsec of the Chandra position, we find two Gaia sources  
listed in the Early Data Release 3 archive \citep{brown21}, but only one is coincident with
the X-ray source detected by Chandra (see Fig. \ref{chandra}), while the other lies too far away from the 
Chandra positional  error circle. The Gaia counterpart of IGR J12134-6015, Gaia 6058696067111698560 at 0.37 arcsec from the Chandra position, 
has coordinates  RA = 12$^{\rm h}$ 13$^{\rm m}$ 23.95$^{\rm s}$ and 
Dec = -60$^{\rm d}$ 15$^{\rm m}$ 16.8$^{\rm s}$ and magnitudes G=18.77  (S/N=13), G$_{\rm BP}$ = 18.01  (S/N=4) and G$_{\rm RP}$ = 17.01 (S/N=5); 
the parallax is 1.471$\pm$0.06 mas, yielding an estimated absolute magnitude of
$\sim$9.6\footnote{To estimate the absolute magnitude we used the 
formula M$_{\rm G}$ = m$_{\rm g}$+5+5Log($\varpi$/1000), where $\varpi$ is the 
source parallax \citep{babusieux18}.} and a distance of of 674 pc \citep{bailer21}, 
indeed suggesting a Galactic source. 

Considering the Gaia colours for this source and following the diagram reported by  \citet{eyer19} 
(see Fig. 2 in their paper), 
IGR J12134-6015 falls in the region just below the Main Sequence occupied by dwarfs and sub-dwarfs. This is also 
supported by  the diagram shown in Fig. 2 of  \citet{abril20}, where again IGR J12134-6015 occupies  a region consistent with both CVs and Dwarf Novae. 
Particularly, IGR J12134-6015 lies in the region where short-period  Polars are found, well inside the range of G$_{\rm 
BP}$-G$_{\rm RP}$ and absolute G magnitude identified for CVs by these authors.
The Gaia counterpart of IGR J12134-6015 is also listed in the ASAS-SN catalog (\citealt{jayasinghe18}
and \citealt{jayasinghe20}) of variable stars, with a parallax consistent with the Gaia measurements, 
again pointing to a Galactic nature for IGR J12134-6015. The optical light curve, found
on the ASAS-SN on-line database\footnote{\tt{https://asas-sn.osu.edu/variables/324039}}, shows an outburst of about 1 magnitude 
at the very beginning of the ASAS-SN coverage, with the amplitude compatible with that of a dwarf nova.

\subsection{Optical/NIR spectroscopy}

Two spectra over the optical/near-infrared (NIR) range were collected at ESO-VLT with XShooter 
(covering the 3000--25000 \AA, range; see 
\citealt{vernet11} for details on the instrument) on Jan 17 and 18, 2019, under the ESO programme 0102.D-0918(A) -- 
PI: S. Chaty -- with exposure times of 556, 1584 and 2479 s in the blue, visual and NIR arms, respectively.

We retrieved the pipeline-analysed, wavelength and flux calibrated spectra for each day from the 
ESO Science Portal\footnote{{\tt http://archive.eso.org/scienceportal/home}}; given that the overall shape was comparable, 
we stacked the spectra together to increase the S/N.

\begin{figure*}
\centering
\includegraphics[scale=0.7]{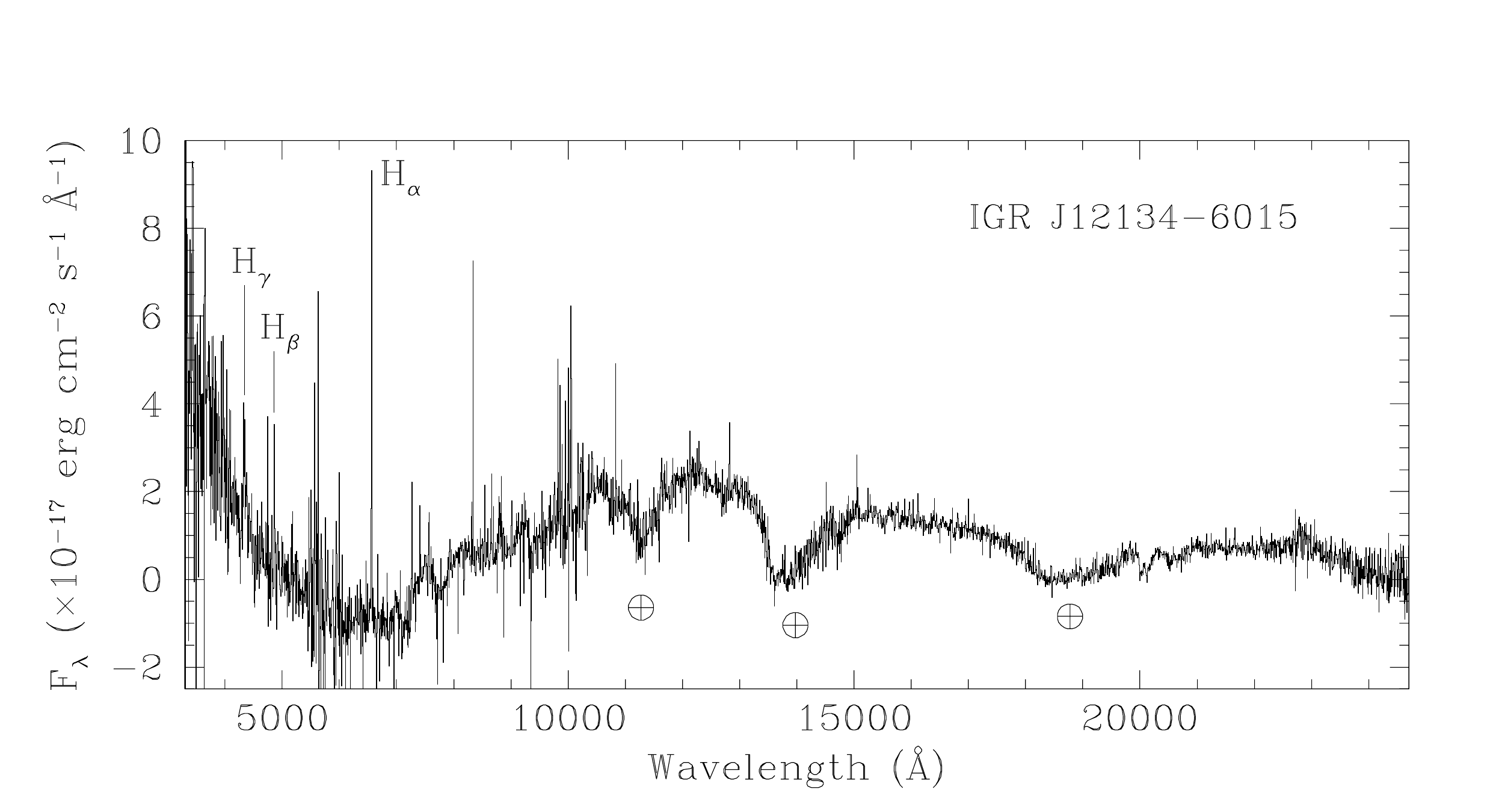}
\caption{{\small Optical-NIR XShooter spectrum of the counterpart of IGR J12134-6015 rebinned at 6 \AA\,($\sim$10 pixels). The NIR telluric features are indicated with the symbol $\oplus$. Apart from H$_\alpha$, all narrow features readily visible in the spectrum are due to noise.}}
\label{spettro}
\end{figure*}

\begin{figure}
\centering
\includegraphics[scale=0.4]{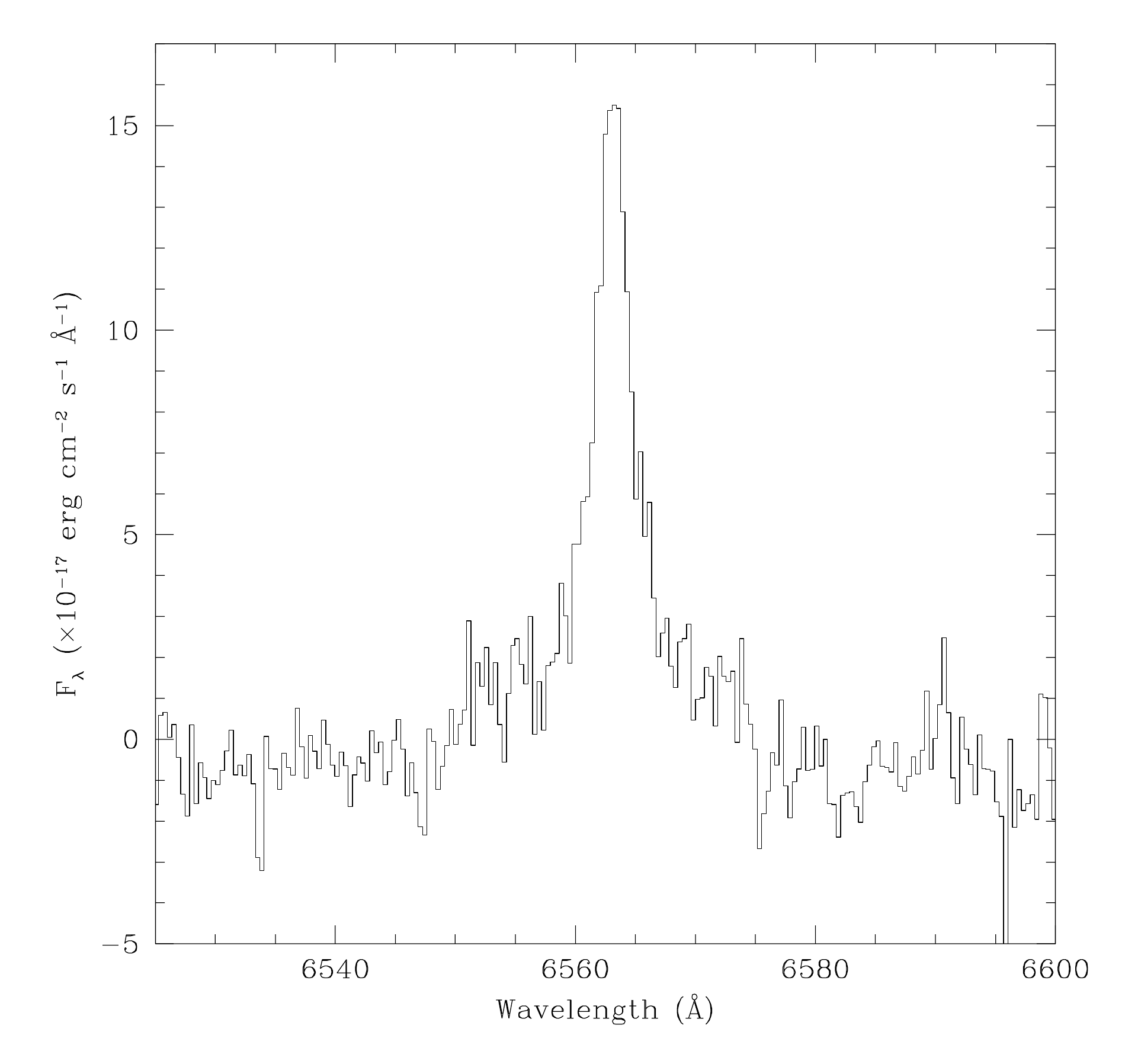}
\caption{{\small Zoom-in on the H$_\alpha$ region of the optical spectrum of IGR J12134-6015.}}
\label{halpha}
\end{figure}

The combined spectrum (Fig. \ref{spettro}) appears to be noisy, with most of the signal at longer (NIR) wavelengths where wide atmospheric bands are apparent, and an increase in flux towards the blue range. Also, H$_\alpha$ and H$_\beta$ lines (and possibly H$_\gamma$) are detected in emission at redshift zero with fluxes (8.5$\pm$0.9)$\times$10$^{-16}$, (2.6$\pm$0.8)$\times$10$^{-16}$  and $\approx$2$\times$10$^{-16}$ erg cm$^{-2}$ s$^{-1}$, respectively. An indication of extended wings is present in the H$_\alpha$ emission profile (see the zoom-in on the
H$_\alpha$ region in Fig. \ref{halpha}), while the S/N is too low in the case of H$_\beta$ to perform a similar investigation. These findings 
confirm that the object belongs to the Galaxy and that it may well be an accreting system composed of a low mass,
late spectral type dwarf 
star (which dominates the emission in the NIR) losing mass onto a compact object with an accretion structure 
around it, possibly a disk, which mostly emits in the bluer parts of the XShooter spectrum. Also, because of the 
detection of short-wavelength photons, the system 
should not suffer from extreme reddening and thus should not be very far from Earth: this consideration on the 
distance is supported by the Gaia data.

If we take the Gaia distance and use the 2-10 keV band flux of 5$\times$ 10$^{-12}$ erg cm$^{-2}$ s$^{-1}$ 
reported by \citet{landi11a} (see also next section), we find that the X-ray output of the source is few times 
10$^{32}$ erg s$^{-1}$, which places IGR J12134-6015 
well within the X-ray luminosity range of CVs observed with INTEGRAL (see e.g. 
\citealt{landi09}. Moreover, as can be seen in Figure 6 of \citet{demartino20}, the source
might occupy the low-luminosity end of the space where this class of objects is located.
Moreover, the optical absolute magnitude of the source determined from the Gaia data is as well 
typical of this kind of source ($\approx$~+9; see e.g. \citealt{warner95}).

\subsection{X-ray analysis}

XRT observed IGR J12134-6015 shortly on four occasions in February 2011, but as the second 
observation was $\sim$200 sec long, we reduced only the longer ones (one on the 6$^{\rm th}$,
which we call Observation 1, one on the 9$^{\rm th}$, which we call Observation 2, and
the last one on 11$^{\rm th}$, called Observation 3; see Table \ref{obs_log}). The
source is not very bright and its detection level is quite low 
(14.7$\sigma$ for Observation 1, 22.4 $\sigma$ for Observation 2 and 12.3$\sigma$ 
for Observation 3). Unfortunately, while there are Chandra HRC-I data available, no data from the ACIS instrument
are present in the archive, so we could not extract a spectrum from the Chandra data.

\emph{X-ray variability}. Since our multiwavelength data strongly suggest that IGR J12134-6015 is a Cataclysmic Variable, 
a class of objects which is characterised by variability both on
short and on long timescales, we 
did a quantitative analysis to see whether variability is present in our X-ray observations
(a thorough timing analysis is  beyond the scope of the present paper). In order to verify if variability
on short timescales (less than an hour) is present, we analysed the XRT light curves in 
each of the three observing periods, using the \texttt{lcstats} tool within the \texttt{ftools}, which 
returns the constant source probability, associated to the Chi-square value, therefore providing
evidence of variability. The analysis of the light curve relative to Observation 1 suggests that
during this period the source has remained more or less constant (with a probability of 
constancy of $\sim$0.82); analysis of the light curves for the subsequent two periods instead points
at some degree of variability, having a probability of constancy of 
0.37$\times$10$^{-17}$ for Observation 2 and of 0.24$\times$10$^{-12}$ for Observation 3. 
Comparing the fluxes of the three observations (see Table \ref{12134_po}), we find that flux
variability (up to a factor of 2) is indeed present on timescales of days. 
From the values reported in Table \ref{12134_po}, it is evident that the source has changed from a low state in the first observation, to a 
high state in the second and then 
returned to a low state in the last one, all in a matter of a few days. 
This is also confirmed by the XMM Slew 0.2-12 
keV flux measured a few years prior (in 2007), where the source was in a low state with a flux of 
2.41$\times$10$^{-12}$ erg cm$^{-2}$ s$^{-1}$  consistent with the source state in  Observation 1 and 3,
but not with Observation 2 where the flux is twice as high as in the other two observations.

In order to assess variability in the hard X-rays,
we ran the task \texttt{lcstats} on the 157-month Crab-weighted BAT light
curve. The statistical analysis indeed suggests that some degree of variability (possibly on
time scales of months) is also seen at higher energies, since the probability
of constancy is 0.38$\times$10$^{-2}$.
Variability is therefore found in a wide range of timescales, as is expected in systems where accretion onto a white dwarf takes 
place; these systems are in fact characterised by variability on different time-scales and across multiple wavelengths.

\emph{X-ray spectral analysis}. As shown above, the source has a certain degree of variability in flux and 
given that also variability in spectral shape cannot be
excluded, we have analysed each of the three observations separately 
and then we fitted the sum of the three spectra, in order to have an average spectrum with higher 
statistics. 

The spectra of the three observations were first fitted with a simple model (\texttt{phabs*po} in
\texttt{XSPEC} terminology) and since the Gaia data suggests that the source is very near, we left the 
column density free to vary, without adding any Galactic N$_{\rm H}$ in order not to  over-estimate the
absorption along the line of sight. This model does not describe sufficiently well the data, since the
reduced $\chi^2$ are 1.54, 2.10 and 0.82 for Observation 1, 2 and 3 respectively (see Table \ref{12134_po}).
However, as can be seen from Table \ref{12134_po}, our spectra are quite poor from a statistical point of view,
except for Observation 2, to allow for the use of more complex models. For this reason,
we attempted to use a more physical model only for data relative to Observation 2. To the simple power-law component,
we added a thermal one, in the form of either a blackbody or a bremsstrahlung (see e.g. \citealt{bernardini12} and \citealt{demartino19}).
From the fits reported in Table \ref{12134_obs2}, it can be seen that the model that best describes the data is the one where the thermal component
is in the form of a bremsstrahlung, whereas the model with the blackbody component returns values for the temperature which are
too high to have any physical meaning and are also not well constrained.
The addition of a power-law component in our fits, although not physically meaningful, is used as an approximation to highlight an underlying spectral complexity at soft energies, which with the current data we are not able to fit with
more appropriate and physical models. We also point out that the fact that the values
for the absorbing column density found in our analyis differ slightly from the
previous values reported by both \citet{landi11} and \citet{karasev12}; this could be due to the fact that while these authors used a summed spectrum of the three observations, here we consider each observation separately and only afterwards the sum of the three. This can lead, due to the poorer statistics of single observations, to larger errors and /or discrepancies in values.
It is therefore evident that in order to have a better understanding of the
X-ray behaviour of this source, high quality spectra are needed, making IGR J12134-6015 the ideal target for future observations
with X-ray facilities such as XMM-Newton or Chandra.

\begin{table}
\begin{center}
\caption{Spectral parameters for IGR J12134-6015. Model employed is \texttt{phabs*po}.}
\vspace{0.2cm}
\begin{tabular}{ccccc}
\hline
{\bf Obs.} & {\bf N$_{\rm \bf H}$}                & {$\bf \Gamma$}        &{\bf F$_{\bf 2-10}$}                     &{$\bf \chi^{\bf 2}$\bf(d.o.f.)}    \\
          &{\bf cm$^{\bf -2}$} &                       & {\bf erg cm$^{\bf -2}$ s$^{\bf -1}$}& \\
           \hline
    1      &$<$0.27$\times$10$^{22}$                      & 1.03$^{+0.44}_{-0.26}$& (4.12$^{+2.73}_{-1.01}$)$\times$10$^{-12}$   & 9.22 (6) \\
    2      &$>$0.82$\times$10$^{22}$                      &0.85$\pm$0.15          & (8.03$^{+1.22}_{-1.18}$)$\times$10$^{-12}$   & 39.83 (19)\\
    3      &$>$0.82$\times$10$^{22}$                      & 1.35$^{+0.71}_{-0.59}$&(4.46$^{+6.84}_{-2.46}$)$\times$10$^{-12}$  & 2.46 (3)\\
\hline
\end{tabular}
\label{12134_po}
\end{center}
\end{table}

\begin{table}
\begin{center}
\caption{Spectral parameters for thermal models of IGR J12134-6015.}
\vspace{0.2cm}
\resizebox{\linewidth}{!}{%
\begin{tabular}{ccccc}
\hline
\multicolumn{5}{c}{{\bf Observation 2, \texttt{phabs*(po+bb)}.}} \\
\hline
{\bf N$_{\rm \bf H}$}                & {$\bf \Gamma$}        & {\bf kT} & {\bf F$_{\bf 2-10}$}   &{$\bf \chi^{\bf 2}$\bf(d.o.f.)}    \\
{\bf cm$^{\bf -2}$} &                 &{\bf keV}      & {\bf erg cm$^{\bf -2}$ s$^{\bf -1}$}& \\
           \hline
 $>$0.82$\times$10$^{22}$  & 2.07$^{+0.95}_{-0.74}$&2.75$^{+35.61}_{-0.93}$&  (8.95$^{+0.24}_{-0.24}$)$\times$10$^{-12}$   & 27.61 (17) \\
 \hline
 \multicolumn{5}{c}{{\bf Observation 2, \texttt{phabs*(po+bremss)}.}} \\
\hline
{\bf N$_{\rm \bf H}$}                & {$\bf \Gamma$}        & {\bf kT} & {\bf F$_{\bf 2-10}$}                     &{$\bf \chi^{\bf 2}$\bf(d.o.f.)}    \\
{\bf cm$^{\bf -2}$} &                 &{\bf keV}      & {\bf erg cm$^{\bf -2}$ s$^{\bf -1}$}& \\
           \hline 
 $>$0.82$\times$10$^{22}$                      &0.74$\pm$0.16  &0.06$^{+0.06}_{-0.04}$        & (8.59$^{+1.54}_{-1.50}$)$\times$10$^{-12}$    &17.81 (17)\\
 \hline
 \multicolumn{5}{c}{{\bf Average Spectrum, \texttt{phabs*(po+bremss)}.}} \\
\hline
{\bf N$_{\rm \bf H}$}                & {$\bf \Gamma$}        & {\bf kT} & {\bf F$_{\bf 2-10}$}                     &{$\bf \chi^{\bf 2}$\bf(d.o.f.)}    \\
{\bf cm$^{\bf -2}$} &                 &{\bf keV}      & {\bf erg cm$^{\bf -2}$ s$^{\bf -1}$}& \\
           \hline 
 $>$0.82$\times$10$^{22}$           &0.79$\pm$0.11  &0.109$^{+0.07}_{-0.04}$        & (6.55$^{+0.87}_{-0.89}$)$\times$10$^{-12}$   &66.97 (36)\\
\hline
\end{tabular}
}
\label{12134_obs2}
\end{center}
\end{table}

\begin{figure}
\centering
\includegraphics[scale=0.33, angle=-90]{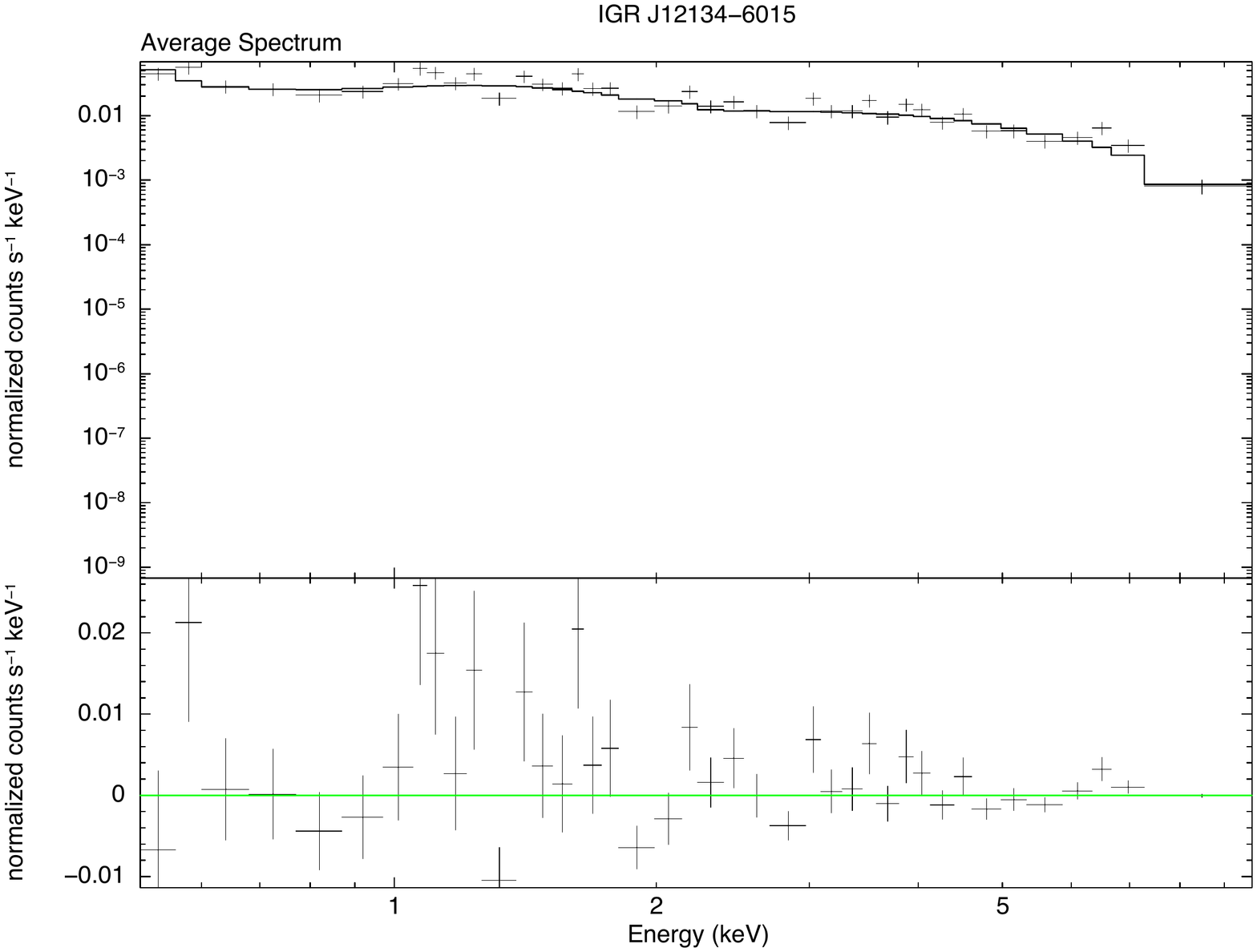}
\caption{{\small 0.5-9.5 keV average spectrum of IGR J12134-6015, the model employed is \tt{phabs*(po+bremss)}}}
\label{12134_average}
\end{figure}

 Since nothing can be deduced from the available spectra regarding spectral variability, we opted to boost the statistics 
of our data by summing the three data sets and fitting the resulting average
spectrum. The results are shown in Table \ref{12134_obs2}; the fit  is not acceptable as evident from the $\chi^2$ value and 
although the bremsstrahlung temperature is well constrained, some residuals are still present
(see Figure \ref{12134_average}). This could be an indication that some degree of spectral 
variability is present in the source, however our data are not statistically good enough to actually
highlight changes in the spectral parameters. For this reason, the summed spectrum must be carefully 
considered and also does not allow to perform a broad-band spectral analysis employing the available INTEGRAL/IBIS
and Swift/BAT spectra, since the high energy data are averaged over very long time periods. 

We can however analyse the high energy spectra from INTEGRAL/IBIS and Swift/BAT and compare them,
in order to assess the source behaviour above 20 keV. We fit the 20-100 keV spectra together using a simple power-law
(leaving the photon index and the normalisation untied for the two spectra)
multiplied by the \texttt{cflux} component so to have an estimate on the fluxes and their errors (see Table \ref{12134_he}). 
Clearly the spectra, which we remind are time-averaged, are very different from one another, a possible indication
of variability even at high energies. In addition, we point out that the INTEGRAL/IBIS spectrum is averaged 
over data taken up to 2013, while the BAT one extends further in time (to about 2018), suggesting that the great 
difference in the two spectra is indeed due to the fact that the source has periods of low and high states which are 
affecting also the high energy, as already highlighted in the timing analysis (see above).

\begin{table}
\begin{center}
\caption{High energy spectral parameters for IGR J12134-6015. Model employed is \texttt{cflux*po}.}
\label{he}
\vspace{0.2cm}
\begin{tabular}{ccc}
\hline
{\bf Instr.} & {\bf$\Gamma$ }& {\bf F$_{\rm \bf 20-100}$}  \\
             &                & {\bf erg cm$^{\rm \bf -2}$s$^{\rm \bf -1}$} \\
           \hline
IBIS        & 2.06$^{+1.00}_{-0.73}$& (5.88$^{+2.29}_{-2.20}$)$\times$10$^{\rm -12}$\\
BAT          & 4.98$^{+0.06}_{-0.11}$ &(3.89$^{+0.16}_{-0.17}$)$\times$10$^{\rm -12}$ \\
\hline
\end{tabular}
\label{12134_he}
\end{center}
\end{table}

Taking into account the evidence provided by our multiwavelength approch, we can conclude
that IGR J12134-6015 is a Galactic object: there is evidence of X-ray variability both 
on short (days) and long (months,  years) timescales, and the optical/NIR and X-ray spectral data analysis, combined with the 
inferred distance, strongly suggests that IGR J12134-6015 is a CV.

\section{IGR J16058-7253}

IGR J16058-7253 was first listed as a high energy emitting source in 2010 in the INTEGRAL 7-year all sky survey by
\citealt{krivonos10}, where it was tentatively associated with the IR source IRAS F15596-7245.
In the BAT 58-month \citep{baumgartner10} and in the 70-month \citep{baumgartner13} catalogues,
the source is associated with the same IR counterpart, but listed as 2MASX J16052330-7253565 and for the
first time a classification as a galaxy is provided. 
The likely counterpart of IGR J16058-7253 was more firmly identified thanks to the Swift/XRT follow-up observation 
performed by \citet{landi11} which detected two sources in the Swift/XRT field of view, located within the 
INTEGRAL/IBIS and Swift/BAT positional uncertainties.
These two X-ray sources were subsequently identified by \citet{masetti13} as two AGN and classified as
LEDA 259580, a Seyfert 2 at z=0.09, and LEDA 259433, a
likely Seyfert 2 at z=0.069 (see in the following).
In the latest INTEGRAL/IBIS survey \citep{bird16}, IGR J16058-7253 was associated to both AGN, which lie just 
3.4 arcmin apart and could not be resolved by IBIS (whose angular resolution is about 12 arcmin) as the
positional accuracy is of the order of 4 arcmin.
However, in the two latest BAT catalogues, the 105-month \citep{Oh18}) and the 157-month
(\url{https://swift.gsfc.nasa.gov/results/bs157mon/}) the likely counterpart of IGR J16058-7253 was
identified as the Seyfert 2 galaxy LEDA 259433. In a subsequent work by \citealt{bar2019}, 
IGR J16058-7253 is classified as an ultra-luminous AGN, characterised by both high bolometric  
(LogL$_{\rm bol}$=45.61 erg s$^{-1}$) and high hard X-ray (LogL$_{\rm 14-195}$=44.71 erg s$^{-1}$) luminosities.

The discrepancy in associations between the BAT
and IBIS catalogues is mainly due to the positions and relative uncertainties derived by the two instruments.
The BAT catalogue reports a position for IGR J16058-7253 at RA = 16$^{\rm h}$05$^{\rm m}$23.28$^{\rm s}$ and
Dec = -72$^{\rm d}$53$^{\rm m}$56.4$^{\rm s}$, with an error circle of 3 arcmin radius; within this error circle, only
LEDA 259433 is found, while LEDA 259580 lies just outside, hence the association proposed in the BAT catalogues.
\citealt{bird16} instead found that the source is positioned at RA = 16$^{\rm h}$05$^{\rm m}$52.8$^{\rm s}$ and
Dec = -72$^{\rm d}$54$^{\rm m}$00$^{\rm s}$, with an error circle of 4 arcmin radius encompassing both sources,
with the two of them correctly listed as likely associations. As a result, at least in the case of the
IBIS detection and possibly also for the BAT one, it is not possible to exclude that both sources 
contribute to the high energy emission above 20 keV.

\begin{figure}
\centering
\includegraphics[scale=0.27]{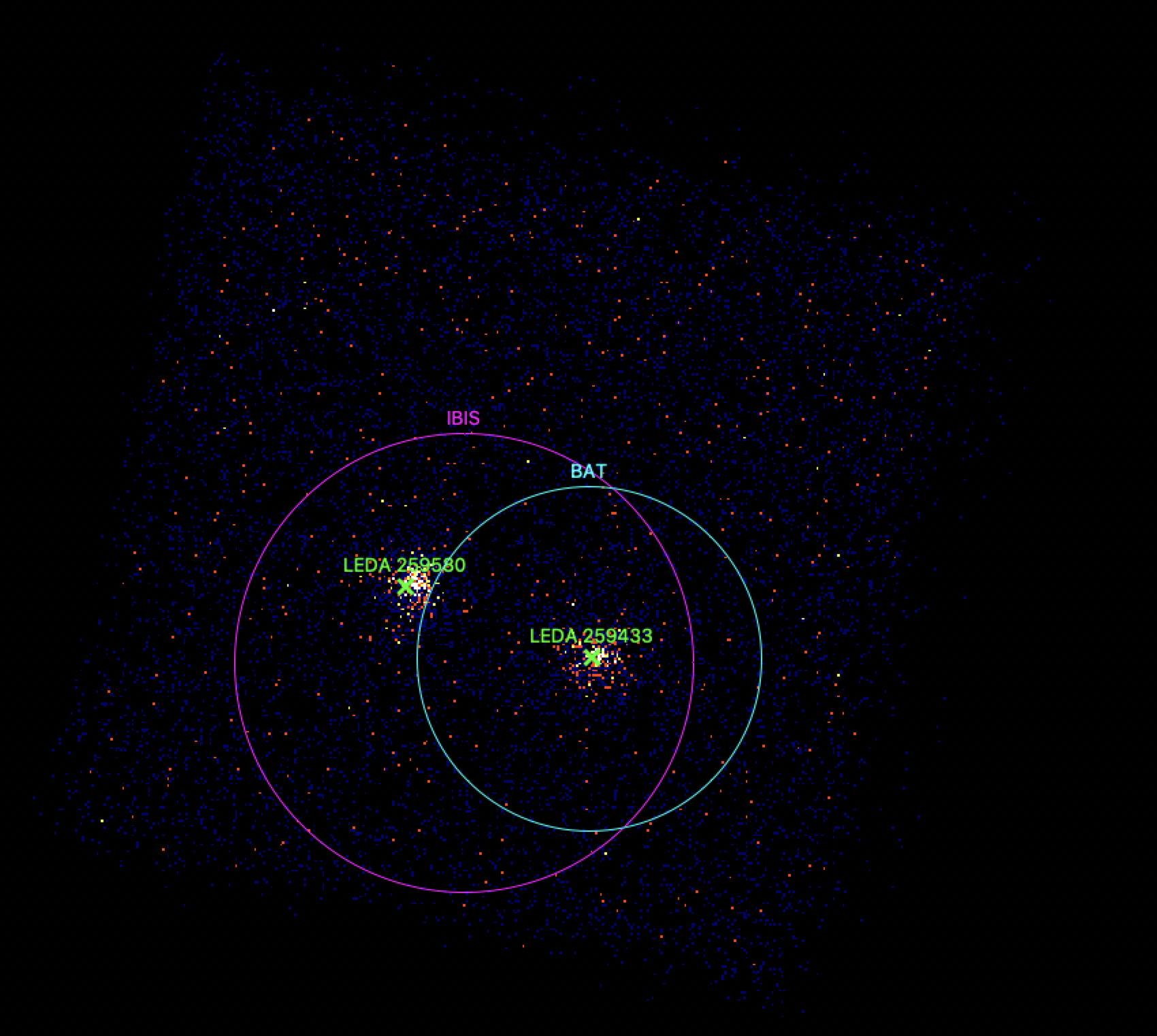}
\caption{{\small 10-80 keV NuSTAR image of the sky region containing the two counterparts of
IGR J16058-7253. The BAT 3 arcmin error circle is shown in cyan, together with the IBIS 4 arcmin error circle
shown in magenta.}}
\label{nu_ima}
\end{figure}

As is often the case with likely counterparts of high energy sources and particularly when an ambiguity is present, 
X-ray data are essential in identifying the correct counterpart. IGR J16058-7253 has been observed by both Swift/XRT
(between 2009 and 2010) and more recently by NuSTAR (in 2019), which, given its imaging capability at high energy, can be valuable in 
determining if only one or both sources emit at energies greater than 10 keV. 
To this aim, we have performed imaging analysis of NuSTAR data by selecting high energy photons ($>$10 keV), and indeed 
inspection of
the 10-80 keV NuSTAR image (see Fig. \ref{nu_ima}), clearly shows that LEDA 259433 and LEDA 259580 are both detected. 

For these reasons, we believe that is not possible to disentangle the
hard X-ray emission from the two sources neither with INTEGRAL, nor with BAT.
Consequently  it is not possible to attribute the measured flux to a single source based on either INTEGRAL or BAT measurements and therefore the estimate of the bolometric and Eddington luminosities for IGR J16058-7253 given by \citet{bar2019} and based solely on the BAT flux is not 
correct.

\subsection{X-ray spectral analysis}
In the following we provide a detailed spectral analysis, by fitting combined XRT/NuSTAR 
spectra of LEDA 259433 and LEDA 259580.

For each source, we fit the XRT and NuSTAR spectra together, employing a
simple phenomenological model, given the poor quality of the data, consisting of a simple power-law
absorbed by Galactic and intrinsic absorption (our baseline model, 
\texttt{const*phabs*phabs*po} in \texttt{XSPEC} terminology).  In the two fits 
we also added a cross-calibration
constant to account for mismatches in the calibration between XRT and NuSTAR and also
to account for flux variability, given the large time-span between the two sets of observations.

\begin{figure}
\centering
\includegraphics[scale=0.33, angle=-90]{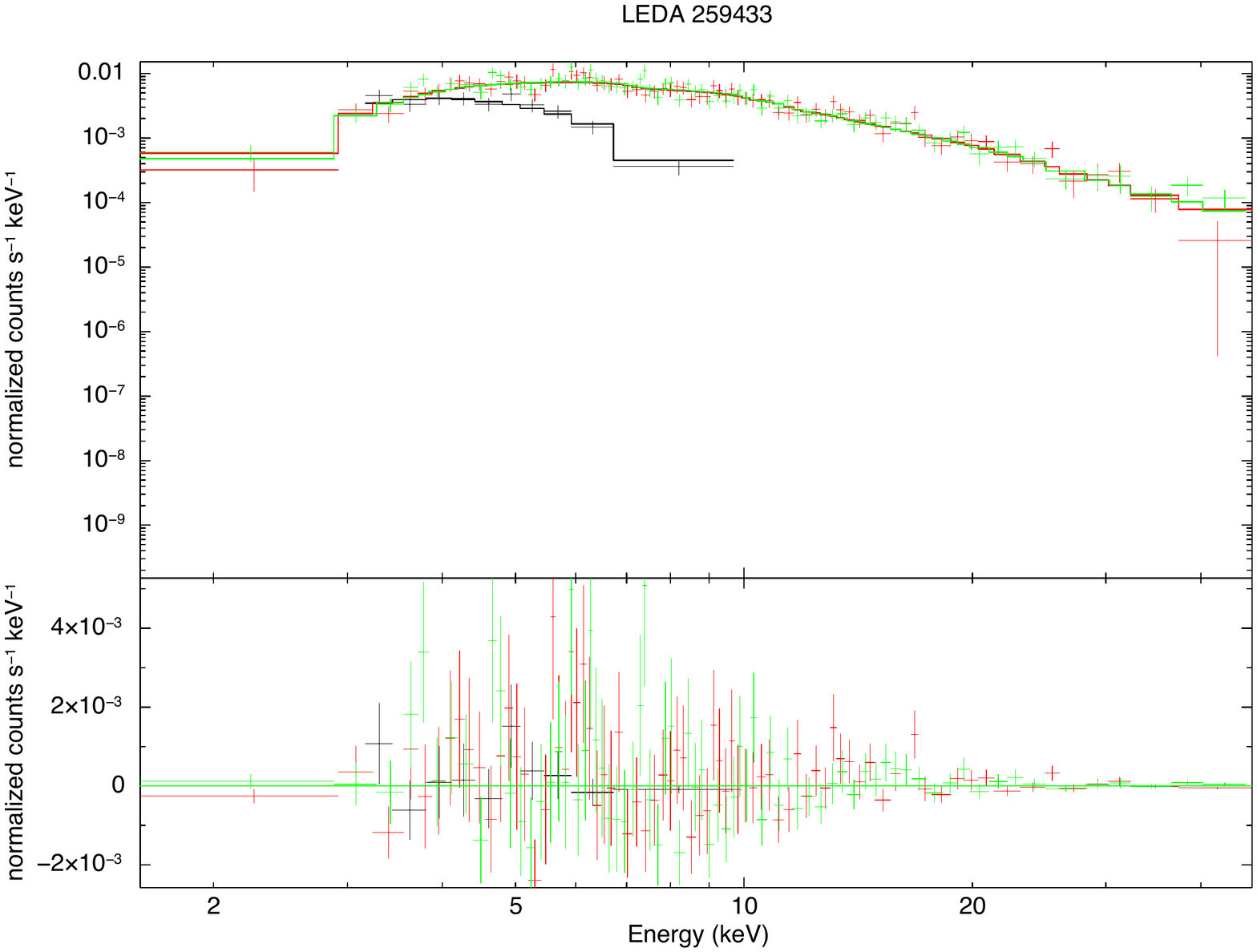}
\caption{{\small XRT/NuSTAR 2-50 keV broad-band spectrum of LEDA 259433. The model employed is \texttt{const*phabs*phabs*po}.}}
\label{leda259433_broad}
\end{figure}

{\it{LEDA 259433}}. The XRT/NuSTAR broad-band spectrum of the likely Seyfert 2 LEDA 259433 covers the 2-50 keV range, 
as there is not enough statistics below 2 keV and above 50 keV to have a spectrum in a broader 
energy range. The data are well fitted
by our baseline model, with a $\chi^2$ of 142.90 for 155 d.o.f.,
resulting in a $\Delta\chi^2$ of 0.92 (see Fig. \ref{leda259433_broad}). 
We find an N$_{\rm H}$ of
(9.43$^{+2.05}_{-1.54}$)$\times$10$^{22}$cm$^{-2}$ (see Table \ref{16058}),
therefore suggesting that the source might indeed be a type 2 AGN as suggested in
previous works (see e.g. \citealt{landi11} and \citealt{masetti13}).
The photon index is
1.56$\pm$0.09, while the cross-calibration constants between XRT and the two NuSTAR detectors are
around 0.9 and 1 respectively. As suggested by the values of the constants, the source 
does not show signs of flux variability, as the 2-10 keV
XRT flux is 2.32$\times$10$^{-12}$ erg cm$^{-2}$s$^{-1}$, while the NuSTAR/FPMA 2-10 keV flux
is 2.05$\times$10$^{-12}$ erg cm$^{-2}$s$^{-1}$. The 20-100 keV flux is 
9.45$\times$10$^{-12}$ erg cm$^{-2}$s$^{-1}$ for NuSTAR/FPMA (the FPMB flux is fully compatible with this value).

\begin{figure}
\centering
\includegraphics[scale=0.33, angle=-90]{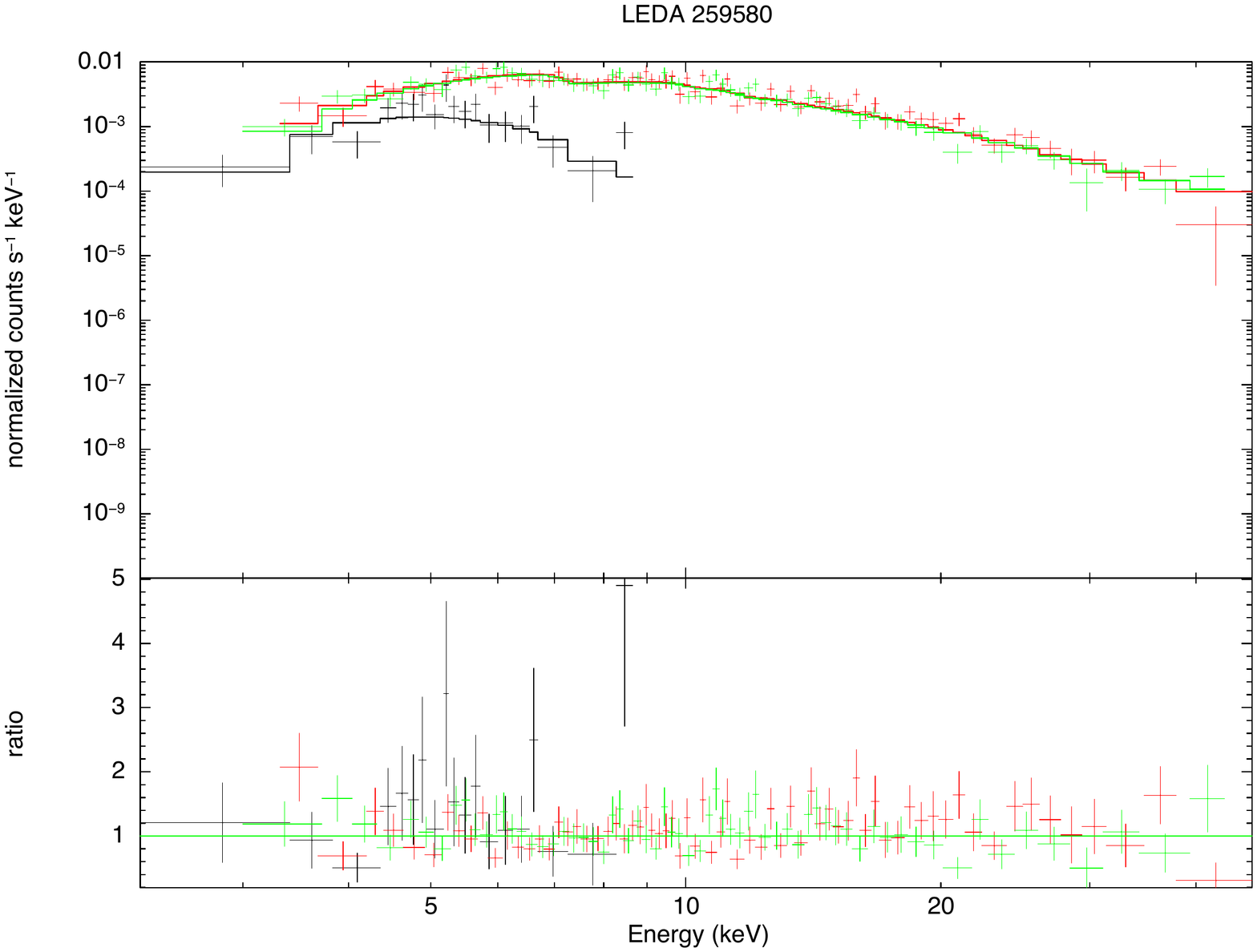}
\caption{{\small XRT/NuSTAR 2-50 keV broad-band spectrum of LEDA 259580. The model employed is \texttt{const*phabs*phabs*po}.}}
\label{leda259580_broad}
\end{figure}

{\it{LEDA 259580}}. As for LEDA 259580, the broad-band fit is performed again 
in the 2-50 keV range, due to the statistical quality of our data. 
The baseline model fits fairly well the data,
having a $\chi^2$ of 171.03 for 149 d.o.f., resulting in a $\Delta\chi^2$ = 1.15 (see
Fig. \ref{leda259580_broad} and Table \ref{16058}).
The intrinsic column density is N$_{\rm H}$=(25.30$^{+4.87}_{-4.31}$)$\times$10$^{22}$cm$^{-2}$,
consisting with the type 2 AGN nature of this source, and
the photon index
is $\Gamma$=1.48$\pm$0.12, while the cross calibration constants are a bit higher
than 1 (around 1.5), suggesting that the source might have undergone some changes in
its flux. Indeed the 2-10 keV XRT flux is found to be 
1.15$\times$10$^{-12}$ erg cm$^{-2}$s$^{-1}$, while the NuSTAR/FPMA flux is 
1.77$\times$10$^{-12}$ erg cm$^{-2}$s$^{-1}$, indicative of minor flux variability. The
20-100 keV flux is instead 1.57$\times$10$^{-11}$ erg cm$^{-2}$s$^{-1}$ for NuSTAR/FPMA
(the value is similar also for the FPMB detector). 

From the fluxes extrapolated from the spectral fits, we can attempt to draw some conclusions on the hard X-ray emission
of these sources. If we take the NuSTAR 20-100 keV fluxes measured from our fits, we find that their sum is 2.51$\times$10$^{-11}$
erg cm$^{-2}$ s$^{-1}$, consistent with the flux obtained from the IBIS spectrum of 2.21$\times$10$^{-11}$
erg cm$^{-2}$ s$^{-1}$, but higher than the flux measured from the BAT spectrum of 1.63$\times$10$^{-11}$ erg cm$^{-2}$ s$^{-1}$. 
From the fluxes reported in Table \ref{16058} it is also evident that LEDA 259580, which was not
identified as one of the possible counterparts of IGR J16058-7253 in the BAT catalogues, appears to be the dominant 
source above 20 keV. Regarding LEDA 259433, from the 2-10 keV flux, we can estimate the luminosity in this band, 
which we found to be 3.14$\times$10$^{43}$ erg s$^{-1}$. From this quantity, we calculated the bolometric luminosity
employing the correction proposed by \citet{marconi04} and we find it to be 7.41$\times$10$^{44}$ erg s$^{-1}$, in 
perfect agreement with the median of bolometric luminosities for type 2 AGN found by
\citet{lusso12}. 
Assuming the black hole mass of 7.08$\times$10$^{7}$M$_{\odot}$ reported by 
\citet{bar2019}\footnote{ BH masses in \citet{bar2019} were derived from the measured velocity dispersions of the Ca H, K and Mg I stellar absorption features, and the employing the relation 
log(M$_{\rm BH}$/M$_{\odot}$)=4.38$\times$log($\sigma_{\ast}$/200 km s$^{-1}$)+8.49 found in \citet{kormendy13}.} for this source, we found that the Eddington luminosity is 8.92$\times$10$^{45}$ erg s$^{-1}$, 
leading to an Eddington ratio of 0.083, again fully consistent with the median value for type 2 AGN reported
by \citet{lusso12}, therefore excluding the possibility of this source being an ultra-luminous AGN.

\begin{table*}
\begin{center}
\caption{Spectral parameters for LEDA 259433 and LEDA 259580. Model employed is \texttt{const*phabs*phabs*po}.}
\begin{tabular}{cccccccc}
\hline
{\bf Source} & {\bf N$_{\rm \bf H}^{\rm \bf gal.}$} & {\bf N$_{\rm \bf H}$}                &{$\bf Gamma$}  &    {\bf F$_{\bf 2-10}^{\rm \bf XRT}$}                             & {\bf F$_{\bf 2-10}^{\rm \bf FPMA}$}                             &{\bf F$_{\bf 20-100}^{\rm \bf FPMA}$}    &{$\bf \chi^{\bf 2}$ \bf{(d.o.f.)}}    \\
             &{\bf cm$^{\bf -2}$}   & {\bf cm$^{\bf -2}$} &             &  {\bf erg cm$^{\bf -2}$ s$^{\bf -1}$}& {\bf erg cm$^{\bf -2}$ s$^{\bf -1}$}& {\bf erg cm$^{\bf -2}$ s$^{\bf -1}$}&\\
           \hline
 LEDA 259433& 0.076$\times$10$^{22}$ (fixed)  & (9.43$^{+2.05}_{-1.54}$)$\times$10$^{22}$ &1.56$\pm$0.09&  (2.32$^{+0.72}_{-0.53}$)$\times$10$^{-12}$ &  (2.05$^{+0.63}_{-0.47}$)$\times$10$^{-12}$ &  (0.94$^{+0.41}_{-0.14}$)$\times$10$^{-11}$& 142.90 (155)\\
 LEDA 259580 & 0.075$\times$10$^{22}$ (fixed) & (25.30$^{+4.87}_{-4.31}$)$\times$10$^{22}$& 1.48$\pm$0.12&  (1.15$^{+0.62}_{-0.41}$)$\times$10$^{-12}$ & (1.77$^{+0.97}_{-0.62}$)$\times$10$^{-12}$ &  (1.57$^{+0.86}_{-0.55}$)$\times$10$^{-11}$ & 171.03 (149)\\
\hline
\end{tabular}

\label{16058}
\end{center}

\end{table*}

\section{Swift J2037.2+4151}
Swift J2037.2+4151 was first reported as a high energy source by \citet{tueller06} and later confirmed in Jem-X pointings
by \citet{westergaard06}. The source was also listed as a transient object in the BAT 58-month
and 70-month catalogues \citep{baumgartner10,baumgartner13}; in the BAT 105-month  \citep{Oh18} and 157-month 
(\url{https://swift.gsfc.nasa.gov/results/bs157mon/}) catalogues,
the source is associated with the IR counterpart SSTSL2 J203705.58+415005.3 and classified as a beamed AGN.
However in \citealt{paliya19}, the source is excluded from the list of BAT blazars on the basis of
its broad-band characteristics. 
Swift J2037.2+4151 was also listed in several IBIS surveys: in the 7-year and
9-year surveys by \citet{krivonos10, krivonos12} and in the latest INTEGRAL/IBIS
catalogue by \citet{bird16}, where it was listed as a transient source; in all these INTEGRAL catalogues, no clear 
classification was given.
The X-ray counterpart of Swift J2037.2+4151 was identified through Swift/XRT follow-up observations 
by \citet{landi11} (see Table \ref{obs_log} for the XRT coordinates and positional error) and was associated
to a 2MASS source, coincident with the NIR counterpart proposed by BAT, which is quite bright and has magnitudes J=16.167$\pm$0.088, H=13.456$\pm$0.033 and K=12.128$\pm$0.020. 
We have investigated other multiwavelength catalogues and found that the IR source is also present in the 
Wide-field Infrared Survey Explorer (WISE, \citealt{wright10}) catalogue, with magnitudes  w$_{\rm 1}$=10.966$\pm$0.026, w$_{\rm 2}$=10.451$\pm$0.024, 
w$_{\rm 3} >$ 10.232 and w$_{\rm 4}>$ 7.896. The WISE colours, w$_{\rm 1} - $w$_{\rm 2}$=0.515$\pm$0.035 and w$_{\rm 2} -$ w$_{\rm 3} <$ 0.219, suggest that Swift J2037.2+4151 is a stellar object, as inferred by the diagram proposed  in Figure 12 of \citet{wright10}. 
This supports the previous indication by \citet{landi11} that Swift J2037.2+4151 is likely a Galactic source, 
either an X-ray binary or a CV.

The optical/NIR counterpart also appears in the Pan-STARRS catalogue \citep{chambers16} with magnitudes $g >$ 26.3, $r >$ 22.4, $i >$ 20.0, $z$ = 21.13$\pm$0.09 and $y$ = 19.51$\pm$0.13.

We moreover performed deep optical $R$-band imaging of the field of Swift J2037.2+4151 on Sep. 16, 2014 with the BFOSC instrument \citep{gualandi01} mounted on the 1.5-m \say{G.D. Cassini} telescope of the INAF-OAS in Loiano (Italy), equipped with a 1300$\times$1340 pixels EEV CCD with a detector scale of 0$\farcs$58 pixel$^{-1}$. The observation started at 22:14 UT; three 20-min exposures were collected under an average seeing of 1$\farcs$8. After standard debiasing and flat-fielding reduction, the images were stacked together; the photometric analysis was carried out using simple aperture photometry and the field was calibrated using USNO-A2.0 stars located close to the XRT position of Swift J2037.2+4151. No source was detected within the soft X-ray error circle of the object down to a 3$\sigma$ magnitude limit R$>$22.5. This limit is consistent with that of the Pan-STARRS catalogue at comparable wavebands.

\subsection{Nature and distance of Swift J2037.2+4151}\label{swift_nature}

Assuming from all of the above that Swift J2037.2+4151 is a Galactic object and an X-ray binary, we can infer its 
nature and distance by placing into context the known multiwavelength information on it.

First, our optical upper limit and the 2MASS near-infrared (NIR) photometry of the source (\citealt{skrutskie06}; see 
also \citealt{landi11}) indicates a very red counterpart for this object, which may at least partially be justified by 
the large absorption along the line of sight detected in the X-rays, as well as by the Galactic reddening ($A_V \sim$ 
16.6 mag according to \citealt{schlafly11}; this figure, by the way, fully explains the non-detection of the source in 
the optical range). However, when we considered the intrinsic NIR colours of stars of different spectral types and 
luminosity classes as tabulated in \citet{ducati01}, we immediately found that no combination of reddening, spectral 
type and luminosity class allows an early-type star as the mass donor in this system: thus, a high-mass X-ray binary 
nature for this object can be ruled out.

Rather, this reddening implies that the object has intrinsic NIR colours consistent with those of an early M giant according to \citet{ducati01}: assuming thus that the object has a M2\,III star as the NIR counterpart, and that its $V$-band absolute magnitude is $-$0.6 \citep{lang92}, we infer that the distance to the source is $d \approx$ 10 kpc, again using the optical-NIR intrinsic colours of stars tabulated in \citet{ducati01}  which imply NIR absolute magnitudes $M_J$ = $-$3.6, $M_H$ = $-$4.4 and $M_K$ = $-$4.5. This distance would place the source within or just beyond the Cygnus arm of the Galaxy according to the map in e.g. \citet{bodaghee12}. Late-type dwarf or supergiant interpretations, albeit having similar intrinsic NIR colours, would return distance estimates of $\approx$100 and $\approx$10$^5$ pc, respectively: both are untenable either due to the large observed absorption, incompatible with a source relatively close to Earth (in the first case), or because of a position which is too deep into the Galactic halo (in the second).
The spectral classification proposed here for the counterpart of Swift J2037.2+4151 may also be tested via the $Q$ parameter diagnostic (\citealt{comeron05}; see also \citealt{negueruela07} and \citealt{reig16}). Following these authors, one can use the 2MASS NIR photometry \cite{skrutskie06} to determine the reddening-free parameter $Q$ = ($J-H$) $-$1.7$\times$($H-K_s$) which, together with the NIR $K_s$ magnitude, allows the construction of a diagram in which early-type and late-type stars occupy different loci: while the latter are mostly concentrated around values  of $Q$ = 0.4–0.5 (which correspond to spectral types K to M), early-type objects typically have $Q \sim$ 0. In the present case, $Q$ = 0.45$\pm$0.11, which places the source right in the range of late-type stars.

This spectral classification is also supported by the WISE and Pan-STARRS data: indeed, using the dereddening coefficients of \citet{wang19}, the intrinsic 2MASS NIR magnitudes and the w1$-$w2 wise colour are compatible with those of a red giant according to \citet{li16}; likewise, the $z-J$ colour of the source, again corrected for Galactic reddening, is similar to that of a K-M late type star (see \citealt{covey07}). This is the most we could extract from the Pan-STARRS and WISE data, given that no information on the $z$, $y$, w1 and w2 intrinsic absolute magnitudes for Galactic stars is readily available in the literature to the best of our knowledge; therefore, no further support to our distance estimate for Swift J2037.2+4151 could be derived from these catalogues.

We also note, as an aside, that the Galactic hydrogen column density N$_{\rm H}$ along the source line of sight is about three times smaller than the one obtained from the dust reddening using the formula of \citet{predehl95}: moreover, no stellar type or luminosity class of any kind, as per \citet{ducati01}, can be recovered by correcting the observed NIR colours with the extinction amount (or lower) associated with the Galactic N$_{\rm H}$. This can therefore be considered just a lower limit on the X-ray absorption towards Swift J2037.2+4151. Indeed, the dust reddening we inferred is more compatible with the N$_{\rm H}$=3.2$\times$10$^{22}$cm$^{-2}$  determined by \citet{landi11} and confirmed by our analysis (see \ref{swift_xray}).

Thus, the distance estimate we infer for Swift J2037.2+4151 implies an X-ray luminosity of $\approx$10$^{36}$ erg 
s$^{-1}$ in the 2--10 keV band. This is at least a couple of orders of magnitude larger than that of hard X-ray 
emitting symbiotic stars (i.e., systems composed by a white dwarf accreting from a red giant star -- see e.g. 
\citealt{smith07}; \citealt{mukai16}; \citealt{danehkar21}). Also, this is a factor $\sim$10$^4$ larger than the typical X-ray output of a CV
(see e.g. \citealt{demartino20}): thus, were this source classified as such, it would lie at a distance of $\sim$100 pc, which (as already stressed above) is too close to justify the large amount of absorption and reddening observed in X-rays and optical/NIR, respectively.
Rather, this X-ray output amount seems to be more typical of Symbiotic X-ray binaries, which, despite the similar name, differ from the above class of objects by the fact that the
accretor is more compact, namely a neutron star, or even a black hole (see \citealt{masetti07} for a sample and the 
main characteristics thereof). This interpretation of course needs a spectroscopic NIR and/or optical confirmation; 
nevertheless, we deem it as the most viable one to account for the amount of multiwavelegth information we examined here.

\subsection{X-ray data analysis}\label{swift_xray}

In the soft X-rays, apart from the JEM-X detection, Swift J2037.2+4151 has been observed three times by Swift/XRT, 
twice in 2006 (in August and December, detected at a significance of 46$\sigma$ and 43$\sigma$, respectively) 
and once in December 2007 (detected at a significance of 22$\sigma$),
while no observations by other X-ray observatories are present in the archives. In the following, we re-analysed the
available XRT data in order to support the classification of Swift J2037.2+4151 based on the multiwavelength 
characteristics discussed in \ref{swift_nature}.

\begin{figure}
\centering
\includegraphics[scale=0.33, angle=-90]{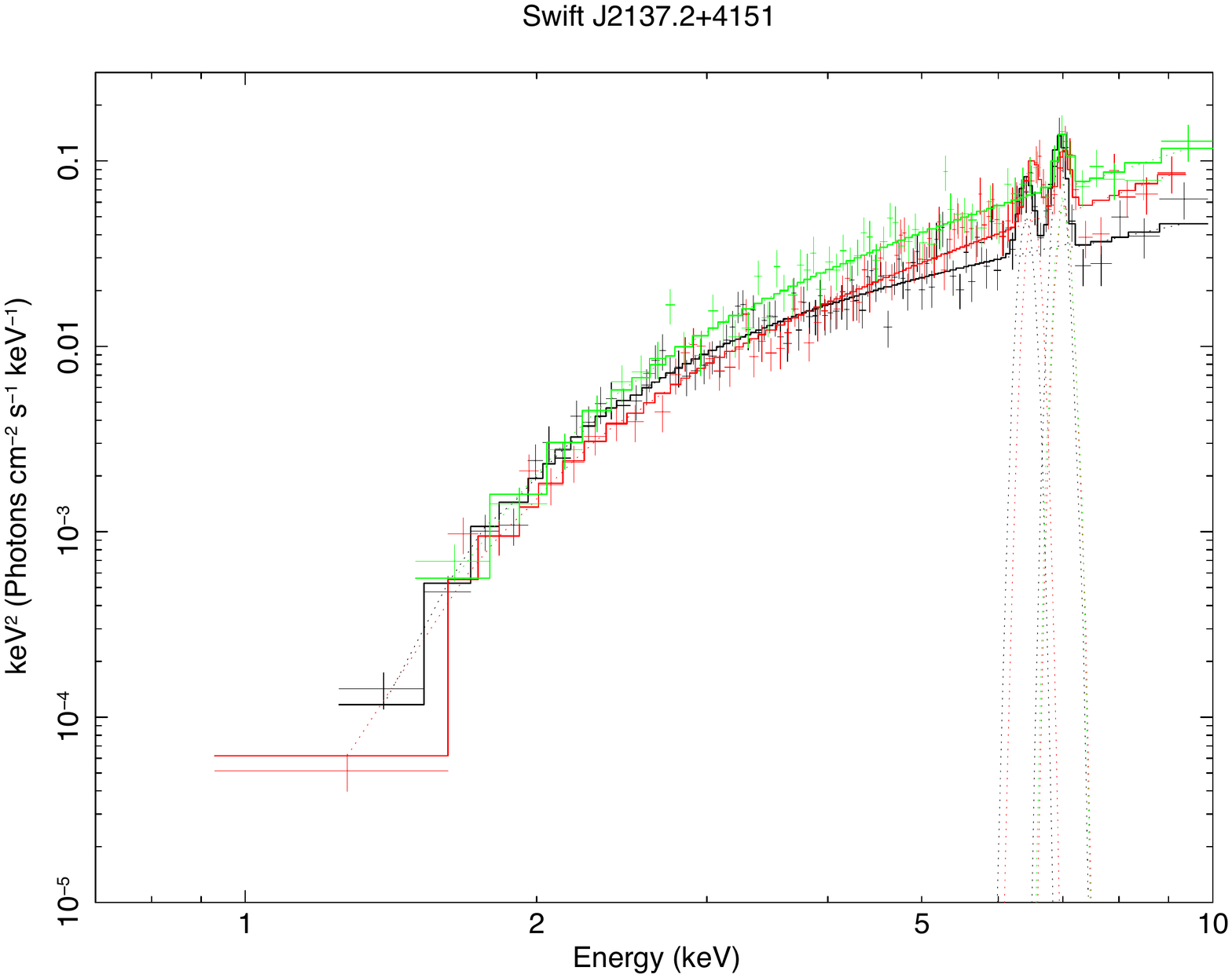}
\caption{{\small  Swift J2037.2+4151 XRT unfolded spectra. Black spectrum: observation 1; green
spectrum: observation 2; red spectrum: observation 3. The model employed is an absorbed power-law plus two
Gaussian components (one in case of Observation 3).}}
\label{swift2037_3obs}
\end{figure}

We fitted the three XRT datasets separately, 
since the source might be variable, not only in flux but 
also in spectral shape in the soft X-rays as well as in the hard X-rays, as expected for Symbiotic X-ray binaries 
(e.g. \citealt{enoto14}). We followed the same approach for each observation; we initially used a very
simple model consisting in a  power-law absorbed by intrinsic N$_{\rm H}$.
As per the discussion  in \ref{swift_nature}, we did not add a Galactic N$_{\rm H}$ in these and subsequent fits, 
but we left the component as a free parameter, in order not to over-estimate the value of the intrinsic column 
density of the source. The simple power-law model does not represent the data
well in any of the observations, since residuals around 6/7 keV, that can be ascribed to the presence of the iron line complex, are
clearly visible in the spectra. Therefore we added a Gaussian component to model the residuals; the component
is required at more than 99.9\% confidence level in all three spectra, but we found that
some residuals are still visible around 7 keV for Observation 1 and 2. We then added
a second Gaussian component in these observations and found that it is required at more than 99.9\% 
confidence level for Observation 1 spectrum and at more than 99.8\% confidence level for Observation 2 spectrum.
The fit results are reported in the upper panel of Table \ref{swift_thermal}, 
and as can be seen by the $\chi^2$ they are all acceptable.
We found that the column density is  compatible with the assumptions made in section \ref{swift_nature}.
The first emission line feature is found at around 6.4/6.5 keV, likely
corresponding to the highly ionised FeXXIV and/or FeXXV. The second
emission line is instead found at $\sim$7 keV and can be
associated with FeXXVI. We point out that the Equivalent Widths (EW) we measured for the Gaussian components 
are slightly larger than what usually found for X-ray binaries (see e.g
\citealt{masetti07} and \citealt{onori21}). This could be due
to the fact that we are likely seeing a blending of several lines, which XRT is unable to resolve; this results 
in a non-physical value of the EW, but nonetheless in line with what is expected in X-ray binaries.
In Figure \ref{swift2037_3obs} we show the unfolded spectra for the three XRT observations relative to this
model.
Despite the fits with the simple power-law are quite good, we also tried a more physical approach to our data by
fitting them with a thermal model in the form of a blackbody (see lower panel of Table \ref{swift_thermal}), but
despite the fits being quite good, we found a blackbody 
temperature which is unusually high for Low Mass X-ray Binaries (see e.g. \citealt{fiocchi07}).

We also checked whether we could apply a 
Comptonisation model to fit our data (\texttt{compTT}) or a \texttt{diskbb} model; these models are 
often used to describe the behaviour of X-ray accreting systems and in particular that of Symbiotic X-ray binaries 
(see \citealt{masetti07} and \citealt{paizis06}). However, both models do not fit the data sufficiently well and do
not allow to put firm constraints on important parameters such as the plasma/disk temperature and optical depth.

From the spectral analysis reported here, it is quite clear that our data lack the statistical significance that
would have allowed us to employ more complex and more physical models. 
Indeed, as is evident from Table \ref{swift_thermal}, the source shows spectral variability, suggesting different
states, which need to be further investigated. Spectra
of higher quality, taken over different time periods, are therefore essential to better characterise 
the spectral behaviour of this source.
Swift J2037.2+4151 also exhibits flux variability in the 2-10 keV band (see Table \ref{swift_thermal}); the source rises steadily from 
Observation 1 through to Observation 3. This
is an expected behaviour in X-ray binaries, which are known to display variability both on short and long timescales (see e.g.
\citealt{corbet08} and \citealt{masetti07}). 

Since the high energy data (INTEGRAL/IBIS and Swift/BAT) are averaged over very long time periods (years), we do not attempt to fit
a broad-band spectrum, since variability could be an issue. 
We therefore analysed data from INTEGRAL/IBIS and Swift/BAT in a similar way as done for
IGR J12134-6015, fitting the 20-100 keV spectra together, using a simple power-law
(leaving the photon index and the normalisation untied for the two spectra),
multiplied by the \texttt{cflux} component, so to have an estimate on the fluxes and their errors. As can be seen
from Table \ref{he_swift}, the two average spectra are consistent within errors, both in spectral shape and flux.

\begin{table*}
\small
\begin{center}
\centering
\caption{Spectral parameters for Swift J2037.2+4151.}
\resizebox{\linewidth}{!}{%
\begin{tabular}{ccccccccccc}
\multicolumn{11}{c}{{{\bf \texttt{phabs*(po+ga+ga)}$^{\star}$}}}\\
\hline
{\bf Obs.} & {\bf N$_{\rm \bf H}$}                &  {$\bf \Gamma$}        &{\bf E$_{\bf line}$}   & {$\bf \sigma$}      & {\bf EW}                  &{\bf E$_{\bf line}$}   & {$\bf \sigma$}      & {\bf EW}                  &{\bf F$_{\bf 2-10}$}&{$\bf\chi^{\bf 2}$ {\bf (d.o.f.)}}    \\
           &{\bf cm$^{\bf -2}$} &                      &{\bf keV}           & {\bf eV}          & {\bf eV}              & {\bf keV}           & {\bf eV}          & {\bf eV}              &{\bf erg cm$^{\bf-2}$ s$^{\bf-1}$}& \\
          \hline
    1      &(3.99$^{+0.61}_{-0.53}$)$\times$10$^{22}$ & 1.07$^{+0.23}_{-0.21}$ &6.41$^{+0.10}_{-0.10}$& 0.10 (fixed) &394$^{+153}_{-98}$& 6.94$^{+0.06}_{-0.06}$& 0.10 (fixed) &737$^{+201}_{-202}$&  (6.48$^{+2.56}_{-2.02}$)$\times$10$^{-11}$  &  87.37 (90) \\
    2      &(3.30$^{+0.63}_{-0.52}$)$\times$10$^{22}$& 0.27$^{+0.20}_{-0.19}$ &6.51$^{+0.13}_{-0.22}$&0.10 (fixed)& 288$^{+139}_{-125}$&7.01$^{+0.10}_{-0.12}$& 0.10 (fixed)& 282$^{+150}_{-158}$&(8.73$^{+3.28}_{-2.20}$)$\times$10$^{-11}$& 88.16 (80)\\
    3      &(3.89$^{+0.98}_{-0.81}$)$\times$10$^{22}$& 0.45$^{+0.26}_{-0.24}$ &6.49$^{+0.17}_{-0.16}$&0.10 (fixed) & 235$^{+175}_{-174}$  & -- &  -- &--  & (1.14$^{+0.65}_{-0.39}$)$\times$10$^{-10}$  &45.74 (57)\\   
\hline
\multicolumn{11}{c}{{{\bf \texttt{phabs*(bb+ga+ga)}$^{\star}$}}}\\
\hline
{\bf Obs.} & {\bf N$_{\rm \bf H}$}                &  {\bf kT}        &{\bf E$_{\bf line}$}   & {$\bf \sigma$}      & {\bf EW}                  &{\bf E$_{\bf line}$}   & {$\bf \sigma$}      & {\bf EW}                  &{\bf F$_{\bf 2-10}$}&{$\bf\chi^{\bf 2}$ {\bf (d.o.f.)}}     \\
           &{\bf cm$^{\bf -2}$} &          {\bf keV}             &{\bf keV}           & {\bf eV}          & {\bf eV}              & {\bf keV}           & {\bf eV}          & {\bf eV}              &{\bf erg cm$^{\bf-2}$ s$^{\bf-1}$}& \\
          \hline
    1      &(2.17$^{+0.38}_{-0.33}$)$\times$10$^{22}$ & 1.87$^{+0.20}_{-0.18}$ &6.43$\pm$0.10& 0.10 (fixed) &392$^{+156}_{-151}$& 6.94$\pm$0.06 & 0.10 (fixed) & 803$\pm$215& (5.97$^{+0.76}_{-0.60}$)$\times$10$^{-11}$  & 98.47 (90) \\
    2      &(2.22$^{+0.42}_{-0.35}$)$\times$10$^{22}$& 2.89$^{+0.43}_{-0.34}$ &6.53$^{+0.12}_{-0.22}$&0.10 (fixed)& 272$^{+135}_{-125}$&7.02$^{+0.10}_{-0.11}$& 0.10 (fixed)& 280$^{+150}_{-158}$&(8.27$^{+2.35}_{-1.56}$)$\times$10$^{-11}$& 85.85 (80)\\
    3      &(2.27$^{+0.62}_{-0.54}$)$\times$10$^{22}$& 2.75$^{+0.47}_{-0.36}$ &6.50$^{+0.15}_{-0.28}$&0.10 (fixed) & 128$^{+120}_{-112}$  & -- &  -- &--  & (1.11$^{+0.39}_{-0.23}$)$\times$10$^{-10}$  &46.55 (57)\\        
    \hline
\end{tabular}
}
\item Notes: $^{\star}$: in the case of Observation 3 data we added only one Gaussian component
\label{swift_thermal}
\end{center}
\end{table*}

\begin{table}
\begin{center}
\caption{High energy spectral parameters for Swift J2037.1+4151. Model employed is \texttt{cflux*po}.}
\label{he}
\vspace{0.2cm}
\begin{tabular}{ccc}
\hline
{\bf Instr.} & {\bf$\Gamma$ }& {\bf F$_{\rm \bf 20-100}$}  \\
             &                & {\bf erg cm$^{\rm \bf -2}$s$^{\rm \bf -1}$} \\
           \hline
IBIS         & 4.76$^{+1.81}_{-1.26}$& (4.71$^{+1.39}_{-1.28}$)$\times$10$^{\rm -12}$\\
BAT          & 5.71$^{+0.34}_{-0.31}$ &(3.94$^{+0.24}_{-0.23}$)$\times$10$^{\rm -12}$ \\
\hline
\end{tabular}
\label{he_swift}
\end{center}
\end{table}

\section{Summary}

In this paper we have investigated the true nature of three high energy sources. Taking advantage of the 
multiwavelength data at our disposal, in particular from the NIR to the X-rays, we were able to 
make robust hypotheses on the classification of the sources we have analysed.

By employing Gaia measurements, together with optical data obtained at the ESO-VLT telescope, we were able to 
determine that the proposed classification for IGR J12134-6015 reported in the 157-month BAT catalogue is
not correct. The source is, in fact, a Galactic object, in particular a Cataclysmic Variable. This is also 
supported by the X-ray spectra, obtained by the XRT telescope on board the Neil Gehrels Swift Observatory, 
which can be described by a thermal model 
typical of CVs;   source flux variability is found, as expected for this class of objects.

As far as IGR J16058-7253 is concerned, we were able to assess, thanks to NuSTAR observations, that the
hard X-ray emission detected by both INTEGRAL/IBIS and Swift/BAT is not coming from just one of the
two counterparts, but rather the high energy detections are the contribution of both AGN. This implies that
IGR J16058-7253, or rather one of its counterparts LEDA 259433, cannot be considered as an ultra-luminous AGN, as suggested by \citet{bar2019}. We were able to estimate spectral parameters for both AGN and, using the
black hole mass considered by \citet{bar2019} for LEDA 259433, we also calculated its bolometric and Eddington 
luminosities, which are well in agreement with the expected values for type 2 AGN, confirming our hypothesis that
this source does not belong to the class of ultra-luminous active galaxies. 

Lastly, multiband analysis of Swift J2037.2+4151 conducted in the optical and X-ray bands, strongly
points to this source being part of the rare and peculiar class of Symbiotic X-ray binaries. From the re-analysis of
the Swift/XRT data we were able to give a general characterisation of the spectral properties of this source; from our
analysis we found that the source is variable both in flux and spectral shape and we also found that Swift J2037.2+4151
exhibits composite features, likely a blending of several emission lines, around the iron line complex.
This supports the assumption that  Swift J2037.2+4151 is likely a Symbiotic X-ray binary; however
more optical, NIR  and good quality X-ray data are needed to further support this hypothesis.

\section*{acknowledgements}

We thank Ivan Bruni and Silvia Galleti of the Loiano Observatory staff for the assistance during the service mode optical observations.

We thank Gianni Catanzaro for useful hints on finding XShooter data and Gisella Clementini for her support in using 
the Gaia archive and data. 

This research has made use of the 2MASS survey catalogues.

This publication makes use of data products from the Wide-field Infrared Survey Explorer, which is a joint project of the University of California, Los Angeles, and the Jet Propulsion Laboratory/California Institute of Technology, funded by the National Aeronautics and Space Administration.

MM acknowledges financial support from ASI/INAF agreement n.2019-35.HH.0 

NM acknowledges the ASI financial/programmatic support via the ASI-INAF agreement n.2017-14-H.0 and the ‘INAF Mainstream’ project on the same subject.

We also thank the anonymous referee for his/her useful comments.

\section*{Data Availability}
The data underlying this work are available in the article. Data not specifically appearing in the article like the some X-ray images and hard X-ray spectra will be shared on reasonable 
request by the corresponding author.

\bibliographystyle{mnras}
\bibliography{biblio}
\label{lastpage}

\end{document}